\documentclass[12pt,a4paper]{article}
\usepackage{color}

\usepackage[english]{babel}
\usepackage{rotating}
\usepackage[utf8x]{inputenc}
\usepackage{amsmath}
\usepackage{graphicx}
\usepackage[a4paper]{geometry}
\usepackage{multirow}

\usepackage{setspace}
\usepackage[labelfont=bf,labelsep=period,justification=raggedright]{caption}

\usepackage{soul}

\makeatletter
\renewcommand{\@biblabel}[1]{\quad#1.}
\makeatother

\date{}

\begin{document}

\begin{flushleft}
{\Large
\textbf{Interface of the polarizable continuum model of solvation with semi-empirical methods in the GAMESS program}
}
\\
Casper Steinmann$^1$,
Kristoffer L. Bl\ae del$^2$,
Anders S. Christensen$^2$,
Jan H. Jensen$^{2,\ast}$
\\
\bf{1} Department of Physics, Chemistry and Pharmacy, University of Southern Denmark, Odense, Denmark
\\
\bf{2} Department of Chemistry, University of Copenhagen, Copenhagen, Denmark
\\
$\ast$ corresponding author, E-mail: jhjensen@chem.ku.dk
\end{flushleft}

\begin{abstract}
An interface between semi-empirical methods and the polarized continuum model (PCM) of solvation successfully implemented into GAMESS following the approach by Chudinov \emph{et~al} (Chem. Phys. 1992, 160, 41). The interface includes energy gradients and is parallelized. For large molecules such as ubiquitin a reasonable speedup (up to a factor of six) is observed for up to 16 cores. The SCF convergence is greatly improved by PCM for proteins compared to the gas phase. 
\end{abstract}

\section*{Introduction}
Continuum solvation models such as the polarized continuum model (PCM) \cite{mennucci2012polarizable} and the conductor-like screening model (COSMO) \cite{klamt1993cosmo} offers a computational efficient model of solvation for molecules treated with electronic structure methods. This paper describes the implementation of an interface between the conductor-PCM (C-PCM) model \cite{klamt1993cosmo,barone1998quantum,cossi2003energies} and the NDDO-based semi-empirical methods implemented in GAMESS \cite{schmidt1993general} (MNDO \cite{dewar1977mndo}, AM1 \cite{dewar1985development}, and PM3 \cite{stewart1989optimization}).  There has been several different implementations of semi-empirical/PCM interfaces \cite{klamt1993cosmo,chudinov1992qm,negre1992new,luque1993am1,caricato2004solvent} and this work follows the implementation proposed by Chudinov \emph{et al}. \cite{chudinov1992qm} However, we also implement the corresponding energy-gradient terms and both the energy and gradient terms are parallelized and tested on relatively large systems such as the protein ubiquitin. 

This paper is organized as follows. 1) We review the relevant expressions for the semi-empirical/PCM interface. 2) We present results of solvation free energies and compare them to previous results. 3) We test the numerical stability for geometry optimizations and vibrational analyses. 4) We present timings and parallelization speed-ups for protein-sized systems. 5) We summarize our findings and provide possible ideas for future improvements. 

\section*{Background and Theory}
In PCM, a molecule (the solute) is placed inside a solvent-cavity usually described by introducing interlocked spheres placed on the atoms of the molecule. The solvent is described as a polarizable continuum with dielectric constant $\varepsilon$. The interaction between the solute and the solvent is described by the apparent surface charges (ASCs). The PCM equations are solved numerically by dividing the surface area up into a finite set of elements called tesserae with a corresponding ASC $q_i$, an area $a_i$ and a position $\vec{r}_i$. There are several implementations of the PCM \cite{tomasi2005quantum} and in this study we focus on the conductor-like PCM (C-PCM) \cite{klamt1993cosmo,barone1998quantum,cossi2003energies}.  For high dielectric solvents such as water C-PCM yields nearly identical results to the more generally applicable integral-equation-formalism PCM (IEF-PCM) \cite{tomasi1999ief} but requires less computational resources.  

For C-PCM the ASCs $\mathbf{q}$ are determined by solving the following matrix-equation
\begin{equation}
\mathbf{C}\mathbf{q}=-\frac{\varepsilon - 1}{\varepsilon}\mathbf{V}.
\end{equation}
where the matrix $\mathbf{C}$ has the elements
\begin{equation}
C_{ij}=\frac{1}{|\vec{r}_j - \vec{r}_i|},\;\;C_{ii}=1.07\sqrt{\frac{4 \pi}{a_i}},
\end{equation}
and $\mathbf{V}$ is the potential of the solute in the solvent for each tessera $i$. The potential $V(i)$ on tessera $i$ is given as
\begin{equation} \label{eqn:potential}
V(i) = \sum_A \left[ \frac{Z_A}{|\vec{r}_A - \vec{r}_i|} - \sum_{\mu,\nu\in A} P_{\mu\nu}V_{\mu\nu}(i) \right],
\end{equation}
where $A$ runs over all nuclei in the solute at position $\vec{r}_A$ carrying a charge $Z_A$. $\mathbf{P}$ is the density matrix of the solute and $V_{\mu\nu}(i)$ are the interaction integrals over basis functions on a tessera $i$ given as
\begin{equation} \label{eqn:vmunui}
V_{\mu\nu}(i) = \bigg \langle \mu \bigg |  \frac{1}{|\vec{r}_A-\vec{r}_i|} \bigg | \nu \bigg \rangle = (s's'|\mu\nu),
\end{equation}
For NDDO methods the right hand side of equation~\ref{eqn:vmunui} is the interaction between a point charge on the surface (represented as $s's'$ in the NDDO approach) and the basis functions of the solute molecule on atom $A$. The $(s's'|\mu\nu)$ integrals needed in equation~\ref{eqn:vmunui} are listed in Table~\ref{tbl:rotatedintegrals} for $s$ and $p$ functions. The integrals are rotated from a local ideal coordinate system onto the molecular coordinate system. The local coordinate system is defined by the distance between the atom $A$ containing the basis functions $\mu\nu$ and the tessera $i$
\begin{align}
\hat{R} &= \frac{\vec{r}_i - \vec{r}_A}{|\vec{r}_A - \vec{r}_i|} = \frac{1}{R} (R_x, R_y, R_z) = (\hat{R}_x, \hat{R}_y, \hat{R}_z), \\
\hat{u} &= \frac{1}{u}(-\hat{R}_y, \hat{R}_x, 0), \\
\hat{w} &= \hat{R} \times \hat{u}.
\end{align}
and the four unique integrals from Table~\ref{tbl:rotatedintegrals} are \cite{dewar1977semiempirical}
\begin{align}
(s's'|ss)  &= \frac{1}{|\vec{r}_A - \vec{r}_i|}, \\
(s's'|sp_\sigma) &= \frac{1}{2} \left[ \frac{1}{|\vec{r}_A - \vec{r}_i|-D_1} - \frac{1}{|\vec{r}_A - \vec{r}_i|+D_1} \right], \\
(s's'|p_\sigma p_\sigma) &= \frac{1}{4} \left[ \frac{1}{|\vec{r}_A - \vec{r}_i| + 2D_2} + \frac{1}{|\vec{r}_A - \vec{r}_i| - 2D_2} + \frac{2}{|\vec{r}_A - \vec{r}_i|} \right],\\
(s's'|p_\pi p_\pi) &= \frac{1}{2} \left[ \frac{1}{\sqrt{|\vec{r}_A - \vec{r}_i|^2+4D^2_2}} + \frac{1}{|\vec{r}_A - \vec{r}_i|} \right].
\end{align}
Here, $D_1$ and $D_2$ are empirical parameters describing charge-separation for the multipoles. They are defined elsewhere. \cite{dewar1977semiempirical} Following Chudinov \emph{et al.} \cite{chudinov1992qm} the density parameters $\alpha_l$ are set to zero in this work and are therefore not shown in the equations.

The electrostatic interaction of the ASCs $\mathbf{q}$ on the surface and the molecule is treated by introducing the following one-electron contribution to the Fock matrix
\begin{equation}
F'_{\mu\nu} = F_{\mu\nu} + V_{\mu\nu},
\end{equation}
where
\begin{equation}
V_{\mu\nu} = -\sum^{N_{ts}}_{i} q_i V_{\mu\nu}(i).
\end{equation}
Finally, the PCM \textcolor{black}{electrostatic interaction free energy} is calculated as
\begin{equation}
G=\frac{1}{2}\mathbf{V}^T \cdot \mathbf{q}.
\end{equation}
Optimization of the molecular geometry in the PCM field requires the derivative of $G$ with respect to an atomic coordinate $A_x$
\begin{equation} \label{eqn:derivativeg}
\frac{\partial G}{\partial A_x}=\frac{\partial \mathbf{V}^T}{\partial A_x} \mathbf{q} + \frac{\varepsilon}{1-\varepsilon}\cdot\frac{1}{2}\mathbf{q}^T \frac{\partial \mathbf{C}}{\partial A_x} \mathbf{q}
\end{equation}
the last term is computed analytically \cite{li2004improving}. The derivative of the potential with respect to an atomic coordinate is done analytically and we give explicit expressions for all terms in Text~S1.

\section*{Methods}
\subsection*{Computational Details}
The semi-empirical/PCM interface was implemented in a locally modified version of GAMESS \cite{schmidt1993general}. The semi-empirical energy and gradient evaluations were allowed to run in parallel but no efforts were made to parallelize the integral evaluation or the assembly of the Fock matrix since the diagonalization is the major computational bottle-neck for large systems. The evaluation of the electrostatic potential (equation~\ref{eqn:potential}) and its derivative (equation~\ref{eqn:derivativeg}) was parallelized. We note that the remaining semi-empirical integral-derivatives in GAMESS is evaluated numerically.

We compared our implementation to that of Chudinov \emph{et al.} for twenty smaller ammonium and oxonium type molecules used in that study. The structures were generated from their SMILES string (see Table~\ref{tbl:ammonium} and Table~\ref{tbl:oxonium}) using Open Babel \cite{openbabel231,o2011open} and optimized in the gas phase and afterwards using the newly implemented code.

Geometry optimizations used a convergence threshold of $5.0\cdot 10^{-4} \mathrm{\,Hartree\,Bohr^{-1}}$ (OPTTOL=5.0E-4 in \$STATPT). To verify the minima, hessians were calculated for all optimized geometries by double difference (NVIB=2 in \$FORCE). When using PCM for geometry optimizations the FIXPVA \cite{su2009fixpva} tessellation scheme was used (MTHALL=4 in \$TESCAV) and the tesserae count for each sphere was set to 60 (NTSALL=60 in \$TESCAV). For solvation free energies the tesserae count was raised to 960 (NTSALL=960 in \$TESCAV) and the GEPOL-GB (Gauss-Bonet) \cite{pascual1987electrostatic} tessellation scheme (MTHALL=1 in \$TESCAV) was used. 

The Mean Absolute Deviations (MADs) of vibrational frequencies between solvated (s) and gas-phase (g) calculations were calculated by
\begin{equation}
\mathrm{MAD}=\frac{1}{n}\sum_{i=1}^{n} \left|f_i^\mathrm{s} - f_i^\mathrm{g}\right|
\end{equation}

We also carried out single point energies and gradients calculations for  Chignolin (PDB: 1UAO), Trypthophan-cage (PDB: 1L2Y), Crambine (PDB: 1CRN), Trypsin Inhibitor (PDB: 5PTI) and Ubiquitin (PDB: 1UBI). The proteins were all protonated with PDB2PQR\cite{dolinsky2004pdb2pqr,dolinsky2007pdb2pqr} and PROPKA\cite{li2005very} at $\mathrm{pH} = 7$ yielding overall charges of -2, 1, 0, 6 and 0 respectively. Either no convergence acceleration, \textcolor{black}{Direct Inversion of the Iterative Subspace \cite{pulay1982improved} (DIIS=.T. in \$SCF) or Second-Order Self Consistent Field \cite{fischer1992general,chaban1997approximate} (SOSCF=.T. in \$SCF) was used}.  In all cases the C-PCM equation was solved iteratively. \cite{li2003continuum} The timings were performed on up to 24 cores on AMD Optirun 6172 shared-memory CPUs. {\color{black}The method is included in the latest release of the GAMESS program.}
\section*{Results}

\subsection*{Electrostatic Solvation Free Energies}
The electrostatic solvation free energies are presented in Tables~\ref{tbl:ammonium} and \ref{tbl:oxonium} for ammonium and oxonium species calculated using PM3/PCM and compared to results published by Chudinov \emph{et al}.\cite{chudinov1992qm} In general, our results underestimate the electrostatic solvation free energy by an average of -1.3 kcal mol$^{-1}$ and -1.9 kcal mol$^{-1}$. \textcolor{black}{The main source of the difference is likely the fact that Chudinov \emph{et al.} uses the original PCM implementation of Miertus, Scrocco and Tomasi \cite{miertuvs1981electrostatic} often referred to a D-PCM) while we use the C-PCM implementation. The solvation free energies from these implementations can differ by several kcal/mol even for neutral molecules \cite{cances1997new}. (While the reference describes a comparison of D-PCM to IEF-PCM, IEF-PCM and C-PCM yield nearly identical solvation free energies for water.)   Another likely source of error is that we use the GEPOL-GB scheme where Chudinov \emph{et al.} uses a more elaborate scheme to reach convergence of the solvation free energies by subdividing the surfaces incrementally.}

\subsection*{Vibrational Frequencies}
To test the numerical accuracy of the PCM gradients we optimized the molecules listed in Tables~\ref{tbl:ammonium} and \ref{tbl:oxonium}. As indicated in Table~\ref{tbl:optfreq} three of the geometry optimizations (A1, O1, and O2) do not converge. A1 can be made to converge by skipping the update of the empirical Hessian matrix (UPHESS=SKIP) but this does not appear to be a general solution to the problem. While some gradient components in these minimizations are quite large the optimizing algorithm eventually settles on a zero step size causing the optimization to effectively stall.  The cause of this behavior is not clear since it is only observed for the smallest systems and was not investigated further. The resulting geometries still lead to a positive definite Hessian and the frequencies are not unusually different from the gas phase values. 

In four cases (A7, O4, O6, O8 and O9) the vibrational analyses yields imaginary frequencies between 26 and 200 cm$^{-1}$.  In the case of O8 and O9 this also occurs for the RHF/STO-3G calculations and in the case of O7-O9 this also occurs for PM3 structures optimized in the gas phase. In most cases the imaginary frequency is associated with the O+ ion and a neighbouring methyl group. The most likely source of these imaginary frequencies is a flat PES associated with the O+ group combined with numerical inaccuracies in the PCM and PM3 gradients.

\subsection*{Timings}
In Table~\ref{tbl:optlarge} we show absolute timings for single point energy and gradient evaluations of proteins either in the gas phase, using DIIS to obtain convergence, or by including the PCM field either with or without SCF convergence acceleration. None of the listed proteins converged in the gas phase without DIIS and even then the SCF converged only for the three smallest proteins: Chignolin, Tryptophan-Cage and Crambine. 

The cost of optimizing the wavefunction in PCM is between two (Crambine) and three (Chignolin and Tryptophan-cage) times more expensive than without. For Chignolin, which is the smallest protein in our test set, it took 21 SCF iterations to converge in PCM while only 13 for PCM/DIIS and 14 for PCM/SOSCF. The other proteins converged within 17 iterations without convergence acceleration and within 14 iterations with. For absolute timings regarding larger proteins, Crambine, Trypsin Inhibitor and Ubiquitin finished in 1293, 3455 and 6732 seconds with PCM without convergence accelleration, but are slower (1314, 3649 and 8777 seconds, respectively) with PCM and DIIS enabled. Using SOSCF did not result in an appreciable decrease in CPU time. \textcolor{black}{The increase in CPU time when using DIIS is due to the extra matrix operations associated with this method, which represent the computational bottleneck for sem-empirical methods.}

Evaluating the ASC potential derivative (equation~\ref{eqn:derivativeg}) analytically has a negligible computational cost compared to evaluating the wavefunction as can be seen from the last column of Table~\ref{tbl:optlarge}.

\begin{figure}
\begin{center}
\includegraphics{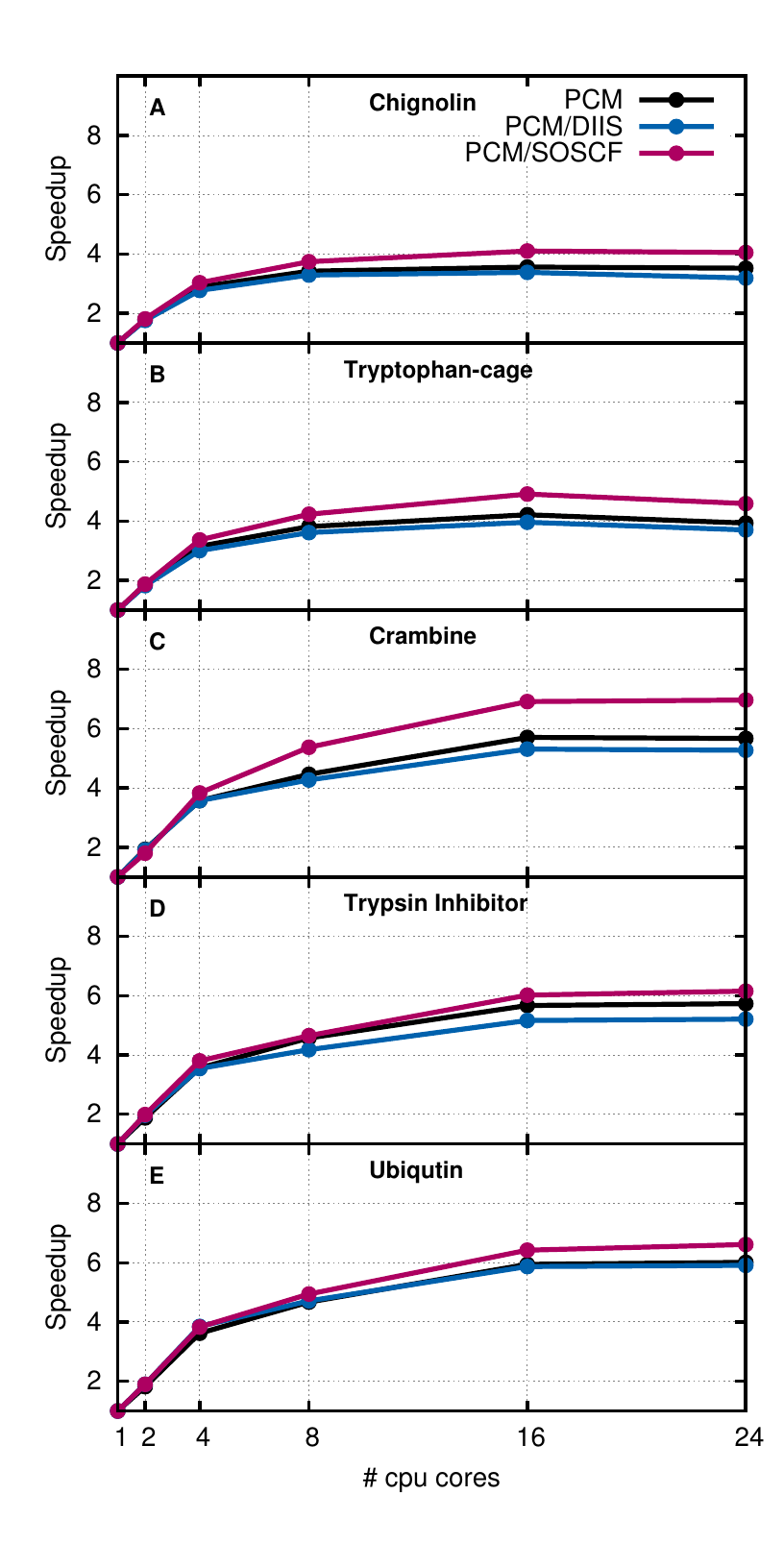}
\caption{Speedup by using multiple cores with PCM enabled for single gradient evaluation.}
\label{fig:pcm_speed}
\end{center}
\end{figure}

The relative speedup from running in parallel in the gas phase is shown on Figure~S1 where no improvement is observed beyond 4 cores (with a speed up factor of 3) and is not discussed further. The PM3/PCM timings (Figure~\ref{fig:pcm_speed}) show better improvement when utilizing multiple cores for all systems. The smaller systems obtain some improvement (a factor 3.4 and 4.2 for Chignolin and Tryptophan-cage, respectively) whereas the larger systems sees improvements of 5.7, 5.7 and 5.9 for Crambine, Trypsin Inhibitor and Ubiquitin, respectively. In all cases maximum speed up is reached for 16 cores because the use of 24 cores introduces some communication overhead which degraded performance.

\section*{Conclusion and Outlook}
An interface between semi-empirical methods and the polarized continuum model (PCM) of solvation successfully implemented into GAMESS following the approach by Chudinov \emph{et~al}. \cite{chudinov1992qm} The interface includes energy gradients and is parallelized.

For very small systems we found some numerical instability problems in the gradient which caused geometry convergence failure, but geometry optimization appears robust for larger molecules.  The use of PCM occasionally introduces imaginary frequencies in the Hessian analysis, but this was also found for RHF/STO-3G PCM calculations and even in a few semi-empirical gas phase calculations so these problems do not appear to be specific to the to the current implementation. \textcolor{black}{We therefore consider the current implementation a working code for all practical purposes, but welcome feedback from readers who encounter numerical stability problems for large molecules} 

For semiemprical methods the most time CPU-intensive part of the calculation remains the solution of the SCF equations.  This part of the code was already parallelized in GAMESS and we show, for the first time, that this implementation applies to semi-empirical methods and the new PCM interface.  For large molecules such as Ubiquitin a reasonable speedup (up to a factor of six) is observed for up to 16 cores.  

It will be interesting to see how much the numerical stability and computational efficiency will improve once the interface is combined with the recently developed FIXSOL/FIXPVA2 method developed by Li and coworkers \cite{thellamurege2012note}.  We are currently working on implementing the PM6 method in GAMESS to further increase the accuracy and range of application that this new interface offers.

\section*{Acknowledgements}
CS would like to thank Lars Bratholm for being tremendous with math. All authors would like to thank Luca De Vico for providing comments on the manuscript.

\bibliography{mopacpcm}

\newpage

\newpage
\begin{sidewaystable}
\begin{center}
\caption{\textbf{Ideal integrals rotated onto the molecular frame.}}
\label{tbl:rotatedintegrals}
\begin{tabular}{ccccc}
      & $s$ & $p_x$ & $p_y$ & $p_z$ \\
$s$   & $(s's'|ss)$ & $(s's'|sp_\sigma)\hat{R}_x$ & $(s's'|sp_\sigma) \hat{R}_y$ & $(s's'|sp_\sigma)\hat{R}_z$ \\
$p_x$ &           & $(s's'|p_\sigma p_\sigma)\hat{R}^2_x + (s's'|p_\pi p_\pi)[\hat{u}^2_x + \hat{w}^2_x]$ & $(s's'|p_\sigma p_\sigma)\hat{R}_x\hat{R}_y + (s's'|p_\pi p_\pi)[\hat{u}_x\hat{u}_y+\hat{w}_x\hat{w}_y]$ & $(s's'|p_\sigma p_\sigma)\hat{R}_x\hat{R}_z + (s's'|p_\pi p_\pi)\hat{w}_x\hat{w}_z$ \\
$p_y$ &           &  & $(s's'|p_\sigma p_\sigma)\hat{R}^2_y + (s's'|p_\pi p_\pi)[\hat{u}^2_y + \hat{w}^2_y]$ & $(s's'|p_\sigma p_\sigma)\hat{R}_y\hat{R}_z + (s's'|p_\pi p_\pi)\hat{w}_y\hat{w}_z$ \\
$p_z$ &           &  &  & $(s's'|p_\sigma p_\sigma)\hat{R}^2_z + (s's'|p_\pi p_\pi)\hat{w}^2_z$
\end{tabular}
\end{center}
\end{sidewaystable}

\newpage
\begin{table}
\begin{center}
\caption{\textbf{Predicted electrostatic solvation free energies of ammonium type molecules.}}
\label{tbl:ammonium}
\begin{tabular}{rcccccc}
               &   & Ref & \multicolumn{2}{c}{PM3/PCM} & \multicolumn{2}{c}{RHF/STO-3G/PCM} \\ \hline
{[}NH4+]       & A1 & 83.9 & 82.4 & (-1.5) & 78.6 & (-3.8) \\
C[NH3+]        & A2 & 73.7 & 72.6 & (-1.1) & 71.3 & (-1.3) \\
CC[NH3+]       & A3 & 70.2 & 69.2 & (-1.0) & 68.6 & (-0.6) \\
CCC[NH3+]      & A4 & 69.9 & 68.5 & (-0.8) & 67.6 & (-1.0) \\
CC([NH3+])C    & A5 & 67.1 & 65.9 & (-1.2) & 66.2 & (0.3) \\
CCCC[NH3+]     & A6 & 69.3 & 68.3 & (-1.0) & 67.1 & (-1.2) \\
CC([NH3+])(C)C & A7 & 64.1 & 62.8 & (-1.3) & 67.1 & (1.2) \\
C[NH2+]C       & A8 & 65.9 & 64.4 & (-1.5) & 65.3 & (0.9) \\
CC[NH2+]CC     & A9 & 59.5 & 58.0 & (-1.5) & 60.7 & (2.7) \\
C[NH+](C)C     & A10 & 59.7 & 57.7 & (-2.1) & 61.8 & (4.2) \\  \hline
AVG & & & -1.3 & &
\end{tabular}
\end{center}
\begin{flushleft}Obtained results using PM3/PCM compared with results by Chudinov \emph{et al}. (labelled "Ref") and RHF/STO-3G/PCM results. PM3/PCM numbers in parenthesis are deviations to the reference. RHF/STO-3G deviations are taken to PM3/PCM results. All numbers are in kcal mol$^{-1}$.
\end{flushleft}
\end{table}

\newpage
\begin{table}
\begin{center}
\caption{\textbf{Predicted electrostatic solvation free energies of oxonium type molecules}}
\label{tbl:oxonium}
\begin{tabular}{rcccccc}
                    & & Ref & \multicolumn{2}{c}{PM3/PCM} & \multicolumn{2}{c}{RHF/STO-3G/PCM} \\ \hline
C[OH2+]             & O1 & 74.1 & 72.6 & (-1.5) & 73.7 & (1.1) \\
CC[OH2+]            & O2 & 69.2 & 67.1 & (-2.1) & 70.2 & (3.0) \\
C[OH+]C             & O3 & 65.1 & 63.4 & (-1.7) & 65.5 & (2.1) \\
C[OH+]CC            & O4 & 61.1 & 59.0 & (-2.1) & 62.5 & (3.5) \\
C1C[OH+]CC1         & O5 & 59.6 & 57.3 & (-2.3) & 61.0 & (3.8) \\
CC[OH+]CC           & O6 & 57.4 & 55.4 & (-2.0) & 59.8 & (4.1) \\
C[OH+]c1ccccc1      & O7 & 54.5 & 53.3 & (-1.2) & 57.4 & (4.4) \\
CC(=[OH+])C         & O8 & 62.5 & 60.0 & (-2.5) & 64.3 & (4.3) \\
CC(C)C(=[OH+])C(C)C & O9 & 53.2 & 51.0 & (-2.2) & 56.0 & (5.0) \\
COC(=[OH+])C & O10 & 60.0 & 58.7 & (-1.3) & 62.6 & (3.9) \\  \hline
AVG & & & -2.0 & &
\end{tabular}
\end{center}
\begin{flushleft}Obtained results using PM3/PCM compared with results by Chudinov \emph{et al}. (labelled "Ref") and RHF/STO-3G/PCM results. PM3/PCM numbers in parenthesis are deviations to the reference. RHF/STO-3G deviations are taken to PM3/PCM results. All numbers are in kcal mol$^{-1}$.
\end{flushleft}
\end{table}

\newpage
\begin{table}
\begin{center}
\caption{\textbf{Optimization steps and frequencies for solvated molecules.}}
\label{tbl:optfreq}
\begin{tabular}{ccccc}
 & \multicolumn{2}{c}{$N_\mathrm{steps}$} & \multicolumn{2}{c}{MAD} [cm$^{-1}$] \\ \hline 
 & PM3 & RHF/STO-3G & PM3 & RHF/STO-3G \\ \hline
A1 & - & 14 & 135.1 & 131.9 \\
A2 & 9 & 10 & 121.5 & 90.8 \\
A3 & 6 & 8 & 64.6 & 39.2 \\
A4 & 6 & 18 & 25.7 & 37.9 \\
A5 & 4 & 17 & 16.9 & 24.6 \\
A6 & 10 & 9 & 30.4 & 15.5 \\
A7 & 32 & 32 & 24.8$^{a}$ & 22.3 \\
A8 & 25 & 24 & 56.2 & 32.8 \\
A9 & 34 & 19 & 27.3 & 31.3 \\
A10 & 32 & 18 & 58.3 & 62.2 \\ \hline
O1 & - & 6 & 151.8 & 60.1 \\
O2 & - & 8 & 111.5 & 36.2 \\
O3 & 15 & 8 & 96.8 & 57.0 \\
O4 & 6 & 8 & 67.1$^{a}$ & 28.3 \\
O5 & 11 & 9 & 85.6 & 29.8 \\
O6 & 15 & 11 & 56.0$^{a}$ & 54.5 \\
O7 & 6 & 6 & 50.1 & 24.5 \\
O8 & 11 & 7 & 87.7$^{a}$ & 22.0$^{a}$ \\
O9 & 6 & 8 & 28.8$^{a}$ & 12.6$^{a}$ \\
O10 & 3 & 6 & 20.8 & 19.9 \\
\end{tabular}
\end{center}
\begin{flushleft}Number of optimization steps for PM3/PCM and RHF/STO-3G/PCM optimizations along with Mean Absolute Deviations (MADs) of vibrational frequencies when going from gas phase to a solvated molecule for all 20 small molecules tested in this work. All optimizations were done in Cartesian coordinates. Translational and rotational frequencies are not included. Dashes marks unconverged structures after 100 optimization steps. $^{a}$ marks optimized structures with at least one imaginary frequency.
\end{flushleft}
\end{table}

\newpage
\begin{sidewaystable}
\begin{center}
\caption{\textbf{Absolute timings for energy and gradient evaluations.}}
\label{tbl:optlarge}
\begin{tabular}{rrccrrrrr}
System            & PDB & $N_{at}$ & $N_{tes}$ & DIIS & PCM & PCM/DIIS & PCM/SOSCF & \\ \hline
Chignolin         & 1UAO & 138 & 1355    & 6 & 29 & 21 & 21 & (0.4) \\
Trp-cage          & 1L2Y & 304 & 2609    & 61 & 159 & 158 & 141 & (1.4) \\
Crambine          & 1CRN & 642 & 4112    & 563 & 1293 & 1314 & 1277 & (6) \\
Trypsin Inhibitor & 5PTI & 892 & 6315    & - & 3455 & 3649 & 3409 & (12) \\
Ubiquitin         & 1UBI & 1231 & 7956   & - & 6732 & 8777 & 7607 & (22) \\
\end{tabular}
\end{center}
\begin{flushleft}Absolute timings in seconds for energy and gradient evaluation for various proteins using both gas phase PM3 and PM3/PCM. Numbers in parenthesis are analytic electronic field gradient timings used in the analytical PM3/PCM gradient. No gas phase SCF converged without DIIS.
\end{flushleft}
\end{sidewaystable}

\newpage
\section*{Supporting Information}
\begin{flushleft}
\textbf{Text S1.} Analytical derivative of the interaction potential.
\end{flushleft}

\end{document}